\documentclass[iop]{emulateapj}

\usepackage{amssymb}
\usepackage{graphicx}
\usepackage{color}    
\usepackage{ulem}

\begin{document}

\title{Reconciling the GRB rate and star formation histories}
\author{Raul Jimenez\altaffilmark{1,2} and Tsvi Piran\altaffilmark{3}}
\altaffiltext{1}{ICREA \& ICC, University of Barcelona (IEEC-UB), Marti i Franques 1, Barcelona 08024, Spain; raul.jimenez@icc.ub.edu}
\altaffiltext{2}{Theory Group, Physics Department, CERN, CH-1211, Geneva 23, Switzerland.}
\altaffiltext{3}{Racah Institute of Physics, The Hebrew University, Jerusalem 91904, Israel; tsvi.piran@huji.ac.il}

\begin{abstract}
While there are numerous indications that GRBs arise from the death of massive stars, the GRB rate  does not follow the global cosmic star formation rate and, within their hosts, GRBs are more concentrated in regions of very high star formation. We explain both puzzles here. 
Using the publicly available VESPA database of SDSS Data Release 7 spectra, we explore a multi-parameter space in galaxy properties, like stellar mass, metallicity, dust etc. to find the sub-set of galaxies that reproduce the recently obtained GRB rate by Wanderman \& Piran (2010). We find that only galaxies with present stellar masses below $< 10^{10}$ M$_{\odot}$ and low metallicity reproduce the observed GRB rate. This is consistent with direct observations of GRB hosts and provides an independent confirmation of the nature of GRB hosts. Because of the significantly larger sample of SDSS galaxies, we compute their correlation function and show that they are anti-biased with respect to the dark matter: they are in filaments and voids. Using recent observations of massive stars in local dwarfs we show how the fact that GRB hosts galaxies are dwarfs can explain the observation that GRBs are more concentrated in regions of high star formation than SNe. Finally we  explain these results using new theoretical advances in the field of star formation.
 \end{abstract}
 
\keywords{galaxies: evolution, galaxies: statistics, galaxies: stellar content, gamma-rays: bursts}

\maketitle

\section{introduction}

Gamma-ray bursts (GRBs) are short and intense pulses of soft gamma-rays. 
It is generally accepted that long-duration ($>2$ s, but see Bromberg et al. 2012)  arise during the Collapse of massive stars,  the so called Collapsars \citep{macwoo}.
This understanding is largely based on the observed association of   several long bursts   with  type Ibc supernovae (SNe) \citep{HjorthBloom12}
\footnote{Most bursts associated with SNe are peculiar low-luminosity GRBs that are produced by a different
mechanism altogether \citep{Bromberg+11}. However, other evidence and in particular 
 the temporal distribution of long GRBs \citep{Bromberg+12} provides a confirmation that long bursts do
 arise during the death of massive stars. } In such a case, it is only natural to expect that GRBs will follow the cosmic star formation rate (SFR)  and that like SNe, their positions within their host galaxies will correspond to the star formation rate within these galaxies. Surprisingly both expectations are not satisfied by observations. 

During the last eight years  the {\it Swift} satellite has routinely provided us with  accurately localized GRBs.  From this data it is possible to construct the luminosity function and cosmic GRB rate. Recently \cite{WandermanPiran} (WP hereafter) have  estimated the rate and luminosity function of long duration  GRBs using a novel method  that solves  simultaneously for the GRB rate and the GRB luminosity function. One of their most interesting findings was that, assuming that the GRB luminosity function does not depend on cosmic time, the GRB event rate does not follow the star-formation rate of the typical galaxy population (their Fig.~9), showing deviations both at low ($z<3$) and large  ($z>3$) redshifts. 
A lot of attention \citep{Daigne+06,LeFloch+06,GuettaPiran07,Guetta+07,Kistler+09,WandermanPiran,Virgili+11,Robertson} was paid to the deviation of the GRB rate from the directly measured SFR  obtained using various methods e.g. the UV luminosities of galaxies at high redshifts \citep{Bouwens+09}. 
While the deviation at high redshifts is clear, both measurements suffer at that redshifts from poor statistics and possible observational biases.  Here we focus on the less noticed but statistically significant and easily verified deviation of the GRB rate from the well documented SFR at low ($z<3$) redshifts. 
At the same time, using HST imaging of GRB and SN host galaxies \citet{Fruchter} showed that while the likelihood to find a SNe in a given position within its host galaxy is linearly proportional to the SFR at that point, GRBs are much more concentrated within the regions of the highest SFR  \citep[see also][]{svensson}. Thus, GRBs don't follow the  star formation neither in time and in space. 

An obvious question that arises is how these two issues can be reconciled with the idea that long GRBs arise from the collapse of  massive stars. A related question is of course  what kind of galaxies host GRBs and whether the SFR within these galaxies
is similar to the cosmic one. 
In an new attempt to  address the first question we ignore all available information on GRB hosts and attack this question using a different approach. 
Our way to answer the question - which galaxies  host GRB  - is to have a complete census of star formation histories of all galaxies in the universe and extract the ones that match the observed GRB rate. Here we do it so by exploiting a large SDSS  local sample of galaxies 
and analyzing in detail their fossil record. This allows us to effectively have a complete sample of galaxies up to $z \sim 3$ and compare the star formation rate in low mass low metallicity galaxies with the inferred GRB rate.

Once we address this question,  we turn to the second question: why are GRBs more concentrated in high SFR rate regions than SNe? To address this issue we make the reasonable assumption that GRBs arise from more massive stars than 
SNe \citep{Ostlin+08}. Following this assumption we compare the positions, within low mass galaxies,  of massive ($> 20 M_\odot$) stars with the positions of less massive ($>9 M_\odot$) stars. Finally, we suggest a simple theoretical argument concerning  star formation that explains this trend. 

\section{Methodology}

The spectra of galaxies encode information about the histories of the  stellar population components, dust, and star formation. Various tools have been developed to extract this information \citep[e.g.,][]{Heavens:1999am,Tojeiro:2007wt} and to compare the resulting extracted information with both extrinsic and intrinsic galaxy properties. The MOPED \citep{Heavens:1999am} algorithm implements the general process of reforming a complex dataset (e.g., a galaxy spectra) into a set of parameters (e.g., star formation rate, metallicity) and parameter combinations, assuming uncorrelated noise, such that the data compression is loss less. 

An easily accessible, robust code, is the VErsatile SPectral Analysis\footnote{http://www-wfau.roe.ac.uk/vespa/} \citep[hereafter VESPA, see][for more details]{Tojeiro:2007wt,Tojeiro:2009kk} package, which recovers star formation and metallicity histories based on the galactic spectra using synthetic stellar-population models. The software recovers histories in adaptive age bins according to the signal-to-noise of the galaxy spectrum on a case by case basis and addresses the age-metallicity relation. Two popular synthetic stellar population models are included in the VESPA output, those of \citet{2003MNRAS.344.1000B}, and the \cite{Maraston:2004em} and \citet[][]{Maraston:2008nn}, which differ in their respective resolutions, and the use of empirical libraries to model the thermally pulsating asymptotic giant branch. Furthermore VESPA corrects for Galactic extinction using the dust  maps of \cite{dustmaps}, and fits for the dust in each galaxy using a dust model with either one or two parameters. 
We exploit the VESPA database to compare the observed GRB rate with that different types of SDSS galaxies. In particular, we address the question: is there is any subset of galaxies from the SDSS whose SFR  matches the observed GRB rate? If so, what are the physical properties of these galaxies?

\section{Results}
\label{results}

The  $\sim10^6$ spectroscopically selected galaxies in this work were drawn from the Sloan Digital Sky Survey Data Release 7 \citep[][and references therein, hereafter SDSS DR7]{York:2000gk,SDSSDR7}.  We selected both ``Main Galaxy Sample" and Luminous Red Galaxies (LRG). Galaxy spectra were reprocessed using VESPA and we followed \cite{Tojeiro:2009kk} in adopting the two parameter dust model, but note that our results are insensitive to the choice of the dust models.

\begin{table}
   \centering
  \begin{tabular}{|  c  |  c |  c  |} 
  \hline
BinID & $T_B+$  Bin Start [Gyrs] & $T_B+$  Bin End [Gyrs] \\ \hline
$0$ & $0.002$ & $0.074$ \\
$1$ & $0.074$ & $0.177$ \\
$2$ & $0.177$ & $0.275$ \\
$3$ & $0.275$ & $0.425$ \\
$4$ & $0.425$ & $0.657$ \\
$5$ & $0.657$ & $1.020$ \\
$6$ & $1.020$ & $1.570$ \\
$7$ & $1.570$ & $2.440$ \\
$8$ & $2.440$ & $3.780$ \\
$9$ & $3.780$ & $5.840$ \\
$10$ & $5.840$ & $7.440$ \\
$11$ & $7.440$ & $8.239$ \\
$12$ & $8.239$ & $9.040$ \\
$13$ & $9.040$ & $10.28$ \\
$14$ & $10.28$ & $11.52$ \\
$15$ & $11.52$ & $13.50$ \\\hline
  \end{tabular}
    \caption{  \label{t2} The 16 time bins within which VESPA
    determines the star formation fraction in the rest frame of the
    galaxy, which corresponds to the time $t=T_B$.}
\end{table}

We  extract  the star formation histories of the galaxies from the VESPA database for the highest resolution output (16 bins) in lookback-time bins as specified in Table~1. VESPA also returns the total stellar mass, both present and at formation, of the galaxy and the metallicity for each bin. We explore the total mass and the metallicity in the VESPA database searching for a match with the WP data (their fig.~2, upper panel). Note that the derived WP GRB rate shows  two characteristic features that do not match the global star formation history of the general population: the peak is shifted to $z\sim 3$ and below that redshift the GRB rate distribution is flatter than the SFR. At high redshift the GRB rate is higher than the SFR rate inferred using
e.g. the UV luminosities of the galaxies \citep[see also ][]{Jakobsson+12}. 
From previous experience, we know that this  behavior is a characteristic of low-mass galaxies \citep{H04}. Fig.~2 in \citet{H04} shows how the shape and peak of the star formation history changes as a function of the galactic mass; this already points out to low-mass galaxies as  promising candidates. This idea is also supported by the non-detection of extreme-redshift GRB host galaxies in very deep HST imaging (Tanvir et al. 2012).
However, in order to find a good match with the WP GRB rate  a cut in metallicity
is also needed. In particular, we find that if we select SDSS galaxies with present stellar masses $< 10^{10}$ M$_{\odot}$ and metallicity $Z < 0.1 Z_{\odot}$, 
we obtain a good match between the GRB rate and the SFR. This means that we need to exclude about half of the dwarfs to obtain such metallicity cut-off 
\citep[see Fig. 6 in][]{pantermet}.  Any other population produces a steeper SFR with redshift and thus cannot be matched to the GRB rate. 

Our reported metallicities correspond to the whole stellar population of the  galaxy, as derived by VESPA. These are different from those typically inferred locally for GRB hosts. For example, recently, \cite{GrahamFruchter12} have measured the metallically of the star forming regions (HII regions) in identified GRB hosts. They measure oxygen abundances from emission lines. Their derived metallicities are somewhat larger than ours (by a factor of 50\%). This is due to the fact that we measure the whole stellar  population, including also the old stellar population in the irregulars\footnote{{\bf MOPED \citep{Heavens:1999am} recovers the metallicity and star formation history from the fossil record as a function of time. The oldest stars will also be the most metal poor as are formed in a more pristine medium. In general, the overall metallicity of the galaxy will increase with time.}}. This lowers the overall metallicity \citep[see the blue line of Fig.~3 in][]{pantermet}. 
So reporting only the metallically of the latest and most recent burst in star formation would bring our metallicity in agreement with the metallicity reported by \cite{GrahamFruchter12}.

\begin{figure}
 \centering
\includegraphics[width=\columnwidth]{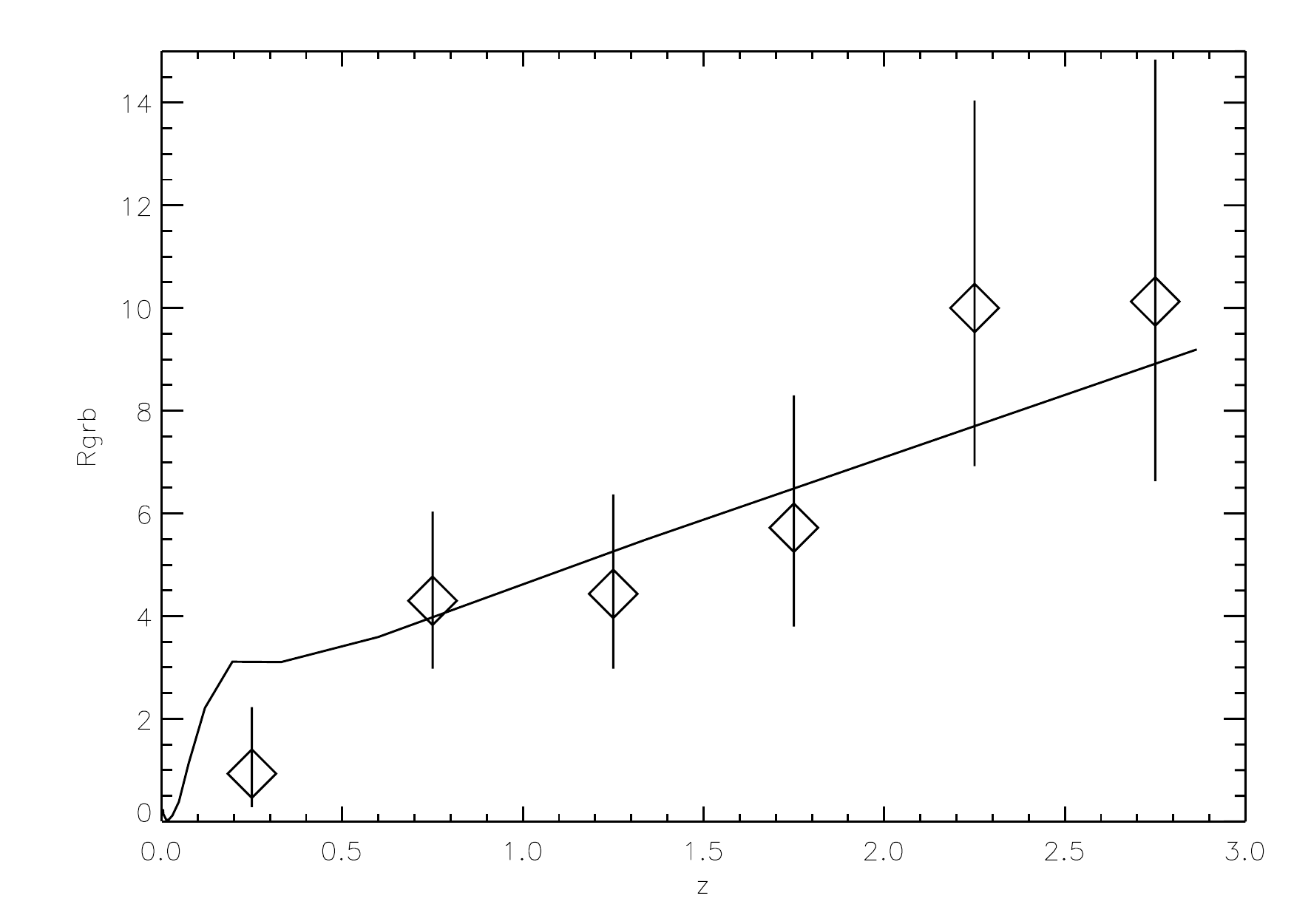}
\caption{The co-moving rate of GRBs from WP (diamonds) with the best fitting star formation history obtained from exploration of the VESPA catalog. Only after imposing a cut-off in galaxy stellar mass ($< 10^{10}$ M$_{\odot}$) and metallicity ($< 1/10 Z_{\odot}$) one finds a good fit (solid line).}  
\end{figure}  

Fig. 1 depicts the WP GRB rate distribution with redshift, where the solid line marks the result of of our fit. First, note that we could not go beyond $z>3$ as the VESPA catalog does not contain information on these high-redshift bins. The reason for this is that we are reconstructing the star formation history from the fossil record at a medium redshift of $z \sim 0.2$ and going back so far in time  is very difficult. Therefore we limit our analysis to redshifts below 3, not covering the peak of the observed GRB rate and the high redshift region beyond it. 

We find that, based just on matching the observed SFR in the specific population of galaxies with the WP GRB rate,   the hosts of long GRBs are dwarf galaxies with very low metallicities. This result is in good agreement with direct observations of those hosts \citep[see e.g.][and subsequent discussion in \S \ref{sec:implications}]{savaglio}, but these samples are limited to a few dozen objects. In fact, the inconsistency between the GRB rate and typical star formation could have been predicted, given that most GRB hosts are low-metallicity dwarf galaxies with a different SFR than the typical star-forming galaxies and hence the overall SFR.
The agreement between the metallicity cut in our analysis and the observational evidence that most GRB hosts have low metallicity further supports the WP results on the GRB rate
and suggests that the specific nature of GRB hosts rather than a luminosity evolution (as suggested by numerous authors)  is responsible for observed distribution of  GRBs redshifts and peak fluxes. 

\begin{figure}
 \centering
\includegraphics[width=\columnwidth]{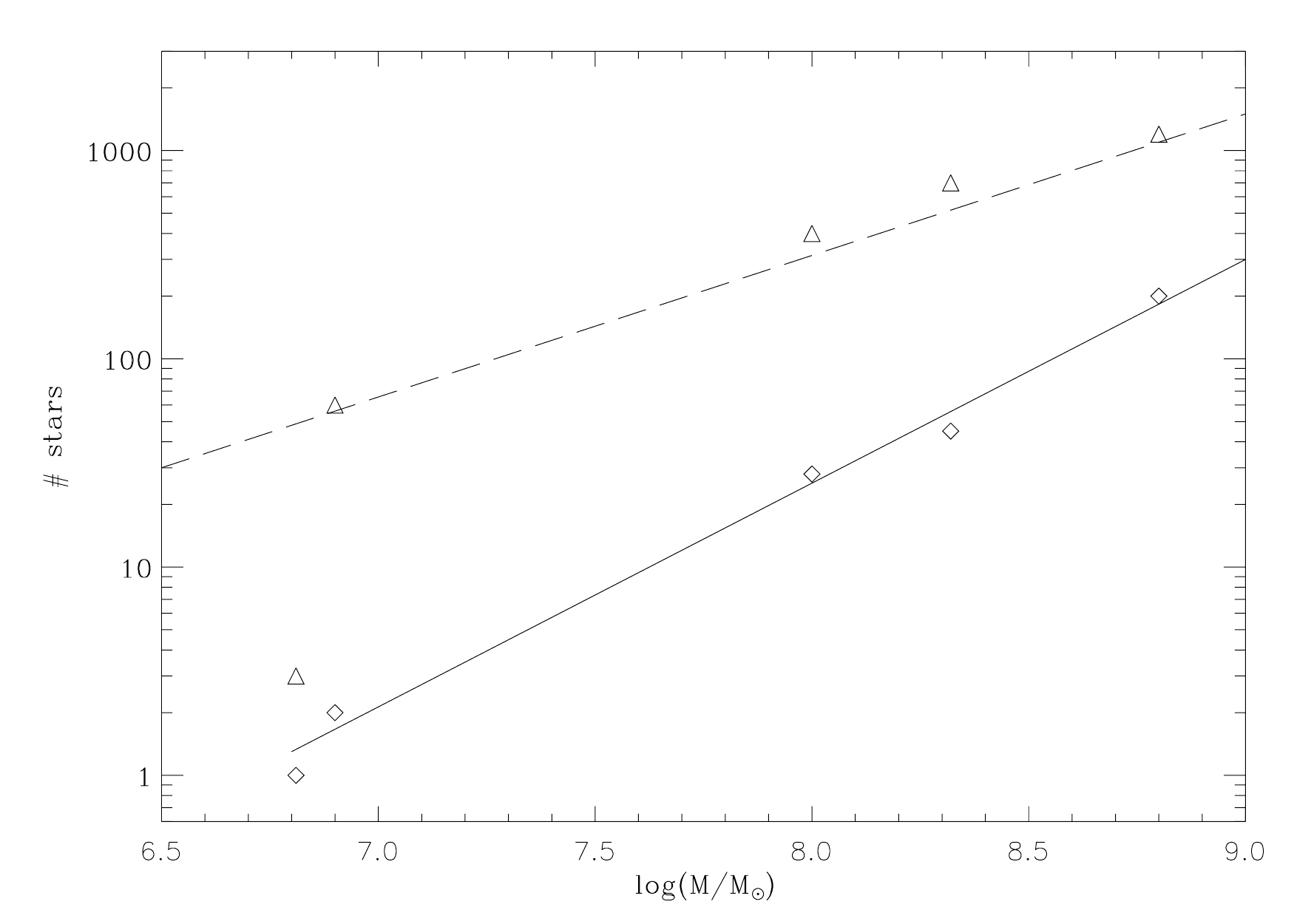}
\caption{Number of stars as a function of the dwarf galaxy mass for five dwarf galaxies. The diamonds correspond to stars with masses $> 20$ M$_{\odot}$ while triangles to $> 9$ M$_{\odot}$. The lines are fits to the data points and show that the more massive dwarfs have more massive stars ($> 20$ M$_{\odot}$) with respect to less massive stars ($> 9$ M$_{\odot}$) than lower mass dwarfs.} 
\label{fig:5}  
\end{figure}

We  now turn to the second puzzle, concerning 
the origin of the discrepancy between the GRB and SN positions within their hosts. To do so we  explore the properties of massive stars in these dwarf low-metallicity galaxies.
Specifically, we compare the population of massive stars  $> 9$ M$_{\odot}$ with the one of very massive stars $> 20$ M$_{\odot}$. The basic idea is that while the former lead to SNe, the latter are  GRB progenitors and if the two populations are distributed differently this explains the different distribution of  GRBs vs. SNe.   

The most detailed study of the population of massive stars in dwarfs in the local group is that by \cite{Bianchi}, who used GALEX observations to determine the number of massive stars ($> 9$ M$_{\odot}$) in six dwarf galaxies (Phoenix, Pegasus, Sextans A and B, WLM and NGC6822), although for the last one they only have lower limits. We have used the masses for these dwarfs derived by \cite{Kara}  (column 4 in their Table~4) to plot in Fig~\ref{fig:5} the numbers of massive stars as a function of the dwarf galaxy mass.  Inspection of  Fig.~\ref{fig:5} reveals that there is a clear correlation between the number of massive stars and the mass of the dwarf galaxy. 
Additionally, the ratio ${(\#> 20)}/{(\#>9)}$ of number of very massive to massive stars increases with the mass of the dwarf galaxy. The lines show fits to the data points in \cite{Bianchi} and are well fitted by power laws with power 0.7 (for $> 9$ M$_{\odot}$) and 1.2 (for $> 20$ M$_{\odot}$), i.e. there is  nearly a factor two difference in the power. This empirical observation implies that very massive stars ($> 20$ M$_{\odot}$) are more abundant in more massive dwarfs. This plot can be compared directly with Fig.~4 in \citet{svensson}, which plots the same quantity but for number of SN and GRBs in hosts galaxies. Our results agree very well with the \citet{svensson} trend.  Given that larger dwarf galaxies are denser than lower mass ones, we can take this relation as a proxy to the relation according to which very massive stars are more abundant  in denser regions in such galaxies. This assertion, that we will shortly explore further, serves now as the basis for the rest of the analysis.

\begin{figure}
 \centering
\includegraphics[width=\columnwidth]{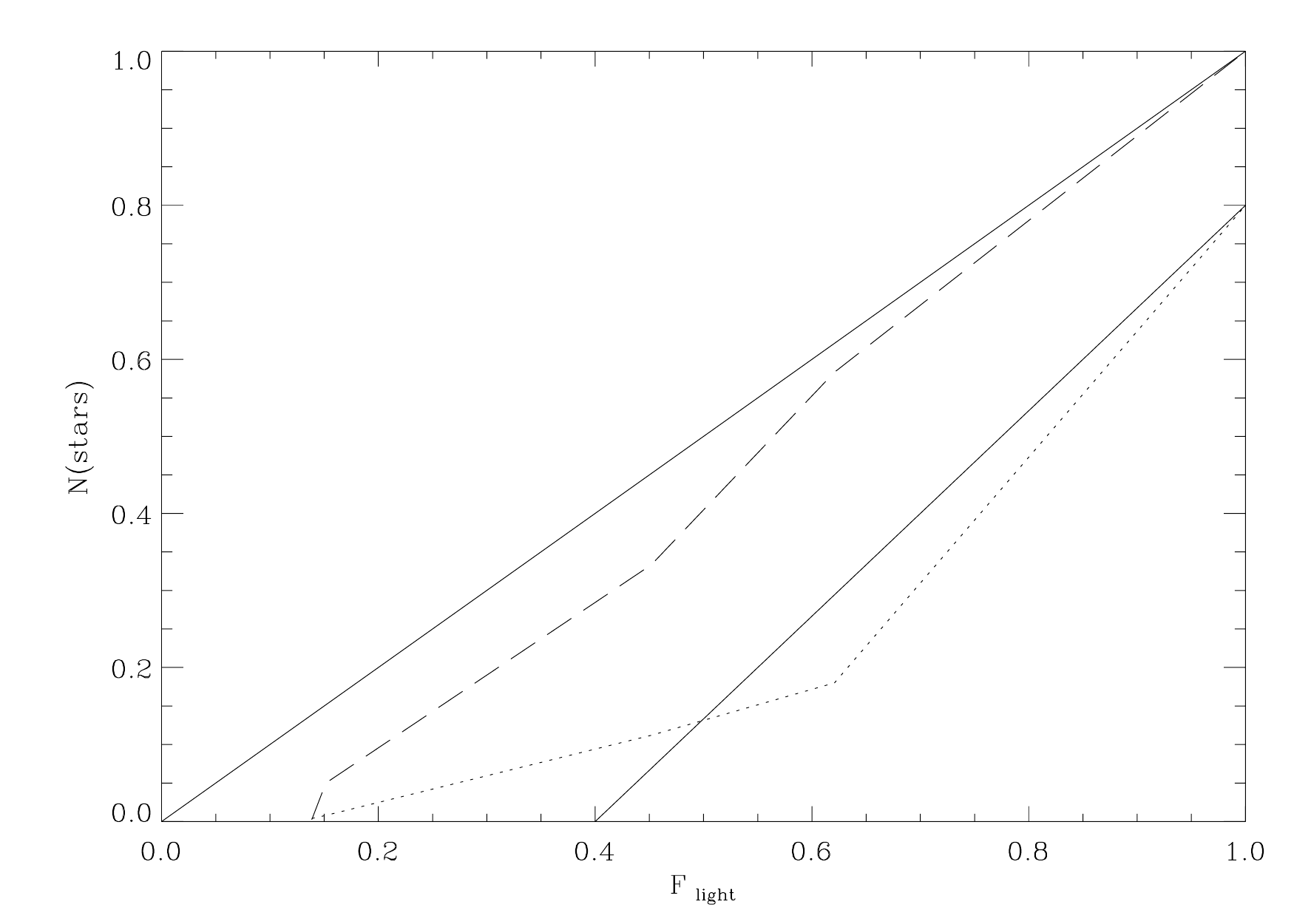}
\caption{Number of stars with mass $> 9$ M$_{\odot}$ (dashed line) or $> 20$ M$_{\odot}$ (dotted line) as a function of the total light in the dwarf galaxy compared with the results from \cite{Fruchter} for the numbers of SNe (upper solid line) and GRBs (lower solid line), note the good agreement.} 
\label{fig:fruchter}  
\end{figure}

We can use the data in \cite{Bianchi} to produce a plot similar to Fig.~3 in \cite{Fruchter}, which shows the number of SN/GRBs as a function of the total light in the galaxy. Because we do not have pixel by pixel information for the dwarfs, we simply integrate the light in radial rings. This is equivalent to what is done in \cite{Fruchter} if the galaxies do not show angular asymmetries, which is the case for the dwarfs we are considering. The results are shown in Fig.~\ref{fig:fruchter}. The solid thick lines are fits to the \cite{Fruchter} results, while the dashed line is the number of stars $> 9$ M$_{\odot}$ and the dotted line corresponds to $> 20$ M$_{\odot}$ star, where we have averaged over the five local dwarfs. There is good agreement between the different curves. 

Because dwarf galaxies have very low surface densities, the outer edges of dwarf galaxies will contain very little mass as compared with the inner part \citep{Tolstoyrev}. For concreteness let us denote as ``outer edges' all mass beyond one equivalent radius and as the  ``inner part' all mass inside it.  For dwarf galaxies this means that there is less than a factor $10$  mass in the outer edges than in the inner part. Using Fig.~\ref{fig:5} this implies that the outer edges will have a factor $3$ less very massive ($> 20$ M$_{\odot}$) stars relative to massive stars ($>9 $M$_{\odot}$) than the inner part. 
If we now assume that GRBs can only originate on very massive stars then this is in agreement with the observations of the spatial distribution of SNe and GRBs by \citet{Fruchter}. Thus the fact that GRBs are mostly hosted in dwarf galaxies, when combined with the dependence of the initial mass function on the density in these galaxies,  explains also the spatial distribution discrepancy between SNe that follow the local SFR linearly and GRBs that follow the local SFR much stronger than linearly.

It is interesting to explore whether there are any theoretical hints that would suggest this observed ``Fruchter et al. law" of a differential initial mass function. The current consensus \citep{sfrev}  is that massive stars form because of the turbulent nature of the interstellar medium. \citet{PadoanNordlund} give a detail theory of star formation due to magnetised turbulence. In particular they predict that the maximal mass of the initial mass function is (their eq.~16): 
\begin{equation}
m_{\rm max} \approx \frac{\rho L^3}{{\cal M}_A} ,
\label{eq:mach}
\end{equation} 
where $\rho$ is the density in the star forming region, $L$ the size of the largest scale on which  turbulence is driven and ${\cal M}_A$ the Alfven Mach number.
It is clear from the above equation that lower density, implies a lower maximal mass. This is the case for dwarf galaxies, in which the density beyond the equivalent radius, is a factor of a few below that in the inner part \citep{Tolstoyrev}. The size of the largest turbulent scale will remain the same while the ${\cal M}_A$ will be similar or slightly larger because of the lower density.
Thus, the maximal mass in the outer lower mass region will be lower as observed (see  Fig.~\ref{fig:5}). Note also that the same authors predict that at lower densities the peak of the mass function will move to larger masses, thus increasing the number of SN.

While the above theoretical argument and the observations of massive stars in local dwarf galaxies would suggest that the number of GRBs should equal the  number of SNe as the mass of the host galaxy increases beyond $10^{10}$ M$_{\odot}$, it is the metallicity cut-of that prevents this from happening.

\section{Implications}
\label{sec:implications}
 The first implication of our results is that, 
 low metallically is 
 an important part of the Collapsar model. While some special cases show GRBs in high metallicity regions, low metallicity is clearly an important factor  for most of the population. This has numerous implications on models of GRBs' progenitors and on the operation of their inner engines. 
This result is not new. In fact  ample information is available on the nature of GRB hosts. Inspection of the hosts 
\citep{Fruchter+99,Chary+02,Bloom+02,LeFloch+03,Tanvir+04,Fruchter,Castro+06,savaglio}; \citep[see also][for recent reviews]{Fynbo+12,Levesque13}  reveals that usually they are low mass  irregular galaxies and have low metallicity \citep{Prochaska+04,Sollerman+05,Fruchter,Modjaz+06,Stanek+06,Thone+07,Wiersema+07,Margutti+07,GrahamFruchter12,savaglio,Thone+13}. 
These findings are consistent with theoretical modeling that suggests that low metallicity is essential to produce 
high angular momentum and high-stellar mass needed for GRB progenitors \citep{YoonLanger05,WoosleyHeger06,WolfPodsiadlowski07}, although these constraints may be avoided by  placing GRB progenitors in binary systems \citep{Fryer,Podsiadlowski10} or uncoupling the evolution of the core and atmosphere of the GRB progenitors \citep{Ekstrom12,Georgy12}.

Recent observations have revealed a population of more massive and dusty GRB hosts \citep{Castro+06,savaglio}, in particular when 
darker (i.e. those with less luminous optical afterglow) GRBs are targeted \citep{Perley+13}.
As the \citet{WandermanPiran} rate concerns only un-obscured GRBs it  is the only one we are able to model. Clearly,  our study only applies to visible GRB hosts and we are unable to exclude the possibility that some GRB hosts have  solar metallicity values (see also \citet{Robertson}).
Nevertheless, it is important to stress that  the fraction of GRBs found in dusty and massive galaxies is not high enough to reconcile the GRB rate with the typical SFR, at least out to z=1 \citep{Perley+13,Kocevski+09}.  Namely, 
given that the fraction of dark GRBs is about 50\% it is clear that obscured GRB hosts do not contribute to the GRB rate more than that  and therefore they cannot modify sufficiently the overall GRB rate and make it compatible with the overall SFR.
Thus, GRBs do prefer galaxies with lower mass with respect to the typical star-forming galaxy population. 
As for the metallicity, while GRB hosts typically have low metallicity, a handful of GRB absorbers show roughly solar abundance \citep{savaglio03,Prochaska+04,savaglio+12} or metal enhancement \citep{DeCia+12}.

Furthermore, \citet{Perley+13} present an extensive compilation of obscured GRBs finding massive and luminous host galaxies at $z > 2$.  But they also point out that at lower redshifts they can only find hosts in low-mass, low-metallicity galaxies. What is the metallicity of a  luminous, massive galaxy a a look-back time of  9-10 Gyr? 
\citet{Perley+13} Fig.~10 shows that the hosts of obscured GRBs have masses in the range $10^9$ to $10^{11}$ M$_{\odot}$. Inspection of Fig. 3 in \citet{pantermet} shows that for the most massive of these galaxies, 9-10 Gyr ago the metallicity of the gas
(i.e. stars formed at that time) will be below solar, with most galaxies in the \citet{Perley+13}  sample below $1/3$ the solar value. \cite{Maiolino} have measured gas metallicities at $z > 2$. Fig. 9 in their paper shows the evolution of gas metallicity as a function of mass. 
Indeed the metallicity is below
solar for all masses at  $z > 2$  and about a dex below solar for most galaxies in the \cite{Perley+13} sample. It seems that the conclusion is that even
for galaxies where the hosts are massive the metallicity of the gas that will form stars is below solar. 

There are, however,  a few systems   
of GRB hosts at high-z ($z > 3$) for which the metallicity has been measured directly , that  show
super-solar values \citep{savaglio+12,DeCia+12}. If it turns out that these are not outliers but represent a common population, then one would have to investigate the GRB host population further. In particular, what is the internal metallicity distribution at high-z in galaxies? It can be very patchy (e.g. \citet{JimenezHaiman}) and how can then one reconcile the observed GRB rate with the SFR inferred from galaxies? 

\begin{figure*}
 \centering
\includegraphics[width=1.55\columnwidth]{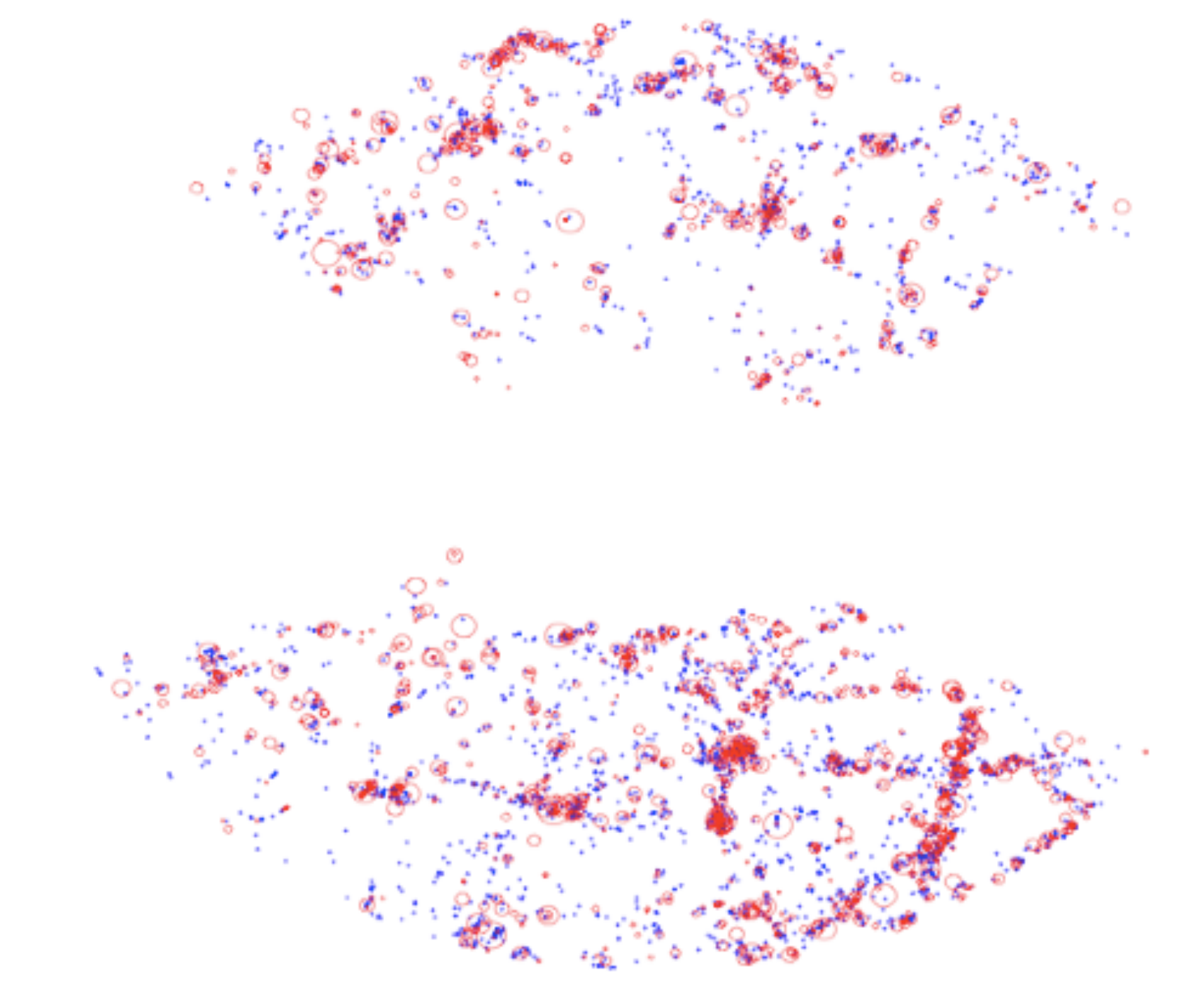}
\caption{Sky location for the host galaxies of visible GRB, that are needed to match the WP GRB rate, is denoted as blue dots. The lower slice corresponds to $z = 0.1$ and the upper slice to $z=3$ (assuming the current location). For reference, we also show galaxies classified as luminous red galaxies, i.e. massive ($> L_{*}$) galaxies with mostly old stellar populations. It is apparent that the hosts of visible GRBs  occupy mostly the filaments and voids of the cosmic web.} 
\label{fig:skypie} 
\end{figure*}

\begin{figure}
 \centering
\includegraphics[width=\columnwidth]{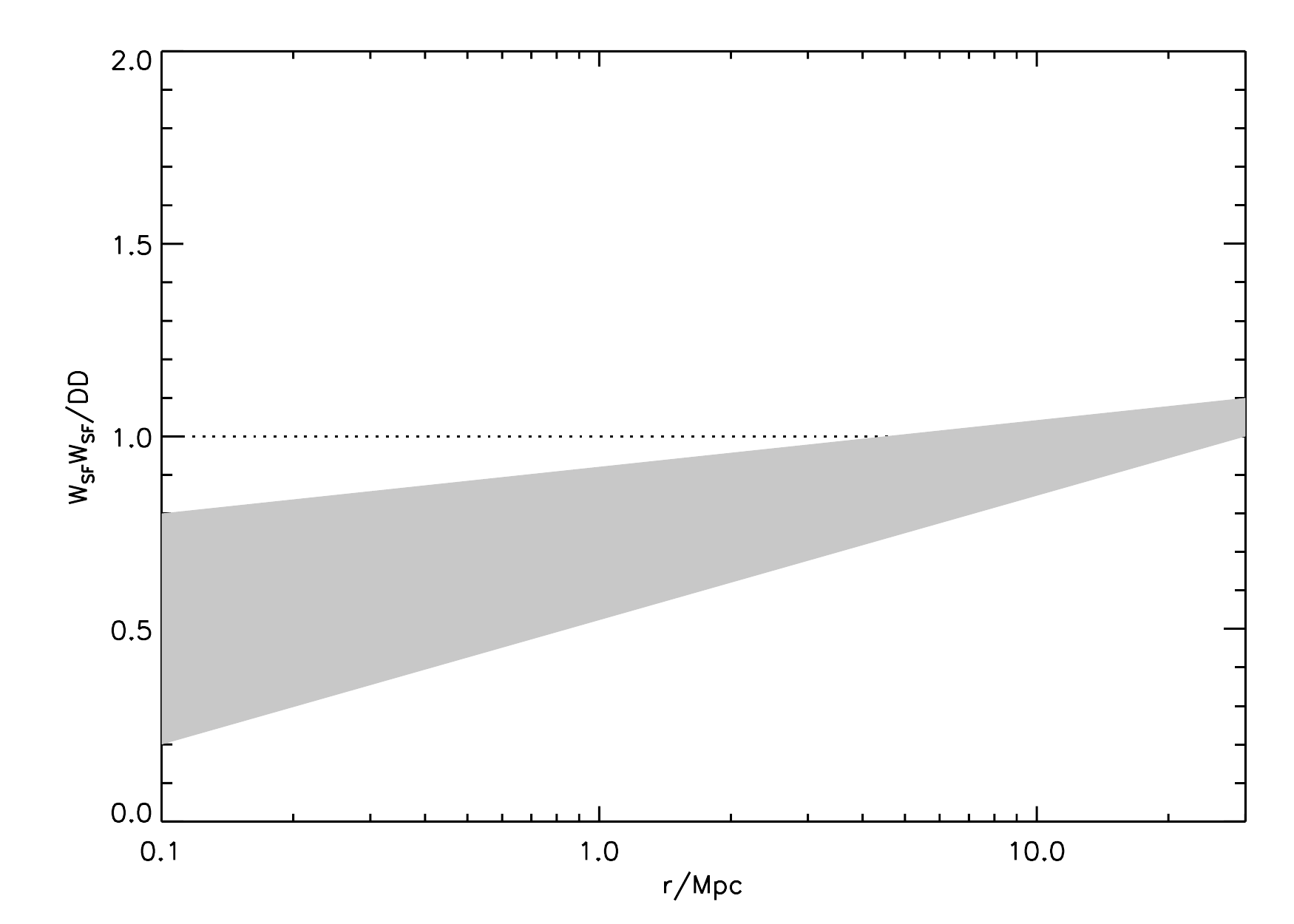}
\caption{Correlation function for the star formation rate in the hosts of visible GRBs as a function of scale. The grey band shows the computed range for the dwarfs in the SDSS DR7 needed to reproduce the GRB rate in WP. Note that GRB hosts are less concentrated than the mean population (dotted line), thus indicating that their pair separation is larger that the mean of the galaxy population.} 
\label{fig:corr} 
\end{figure}  
  
The fact that  low-metallicity plays an important role in the formation of GRBs
 has also far reaching implication concerning the use of GRBs to explore the early universe. Properly utilizing GRBs as probes of the early universe requires a thorough understanding of their formation and the host environments that they sample 
 \citep{Levesque13}. 
Galaxies exhibit a strong mass-metallicity relation \citep[see e.g. Fig.~6 in][]{pantermet} and massive galaxies ($> 10^{11}$ M$_{\odot}$) do not have such low metallicities as those of GRB hosts. Thus, the GRBs host galaxy population  is biased relative to the overall galaxy population and even relative to the population of dwarf galaxies as we had to impose a metallicity cut of $\approx 1/10$ the solar value. This excludes about half of the dwarf galaxies as possible GRB hosts.

This last point suggests that as we move to the early universe, where metallicity was lower, the GRB rate was much higher and a significantly larger fraction of stars resulted in GRBs. This is consistent with the observation in WP that the GRB rate at high redshifts is significantly flatter than the SFR. Note however, that a massive galaxy will enrich its gas extremely fast (in a  dynamical time), thus even if the original gas is of low metallicity, only galaxies that have low rates of star formation will be able to host GRBs as they will be able to keep the gas at low metallicity. As we move to lower redshifts and the overall metallicity of the gas increases, even dwarf galaxies will have difficulties at hosting GRBs if they already have a high metallicity of the available gas to form stars. This explains why the GRB rate decreases at lower redshift. However, because dwarf galaxies dominate the SFR at low-z, the GRB rate decreases slower than the SFR of the overall galaxy population. If we use the metallicity of the damped Lyman$- \alpha$ systems as a tracer for the metallicity of the gas where galaxies form, we see that at redshift  $z \approx 2-3$ the metallicity of the gas decreases below $1/10$ the solar value (see Fig.~12 in \cite{wolfe}). This same conclusion is obtained if one looks at the metallicity histories of SDSS galaxies \citep[see Fig.~3 in][]{pantermet}, thus the GRB rate should increase at that redshift in agreement with WP.

After having identified GRB hosts as low-metallicity dwarf galaxies, we can now explore properties of this population within the 
SDSS data. First,  we examine the location in the sky as a function of RA and declination of this population. Fig.~\ref{fig:skypie} depicts the low metallically dwarf galaxies as blue dots. For reference we have also plotted the locus of luminous red galaxies (i.e. galaxies with larger masses ($\sim L_{*}$) with older  stellar population). It is apparent that the low metallicity dwarfs are located in the filaments-voids of the cosmic web. The bottom panel of Fig.~\ref{fig:skypie} shows the positions at $z=0.1$ while the top panel depicts the position at $z=3$ (for this case we have used the current local positions but identified the galaxies according to their star formation history at $z=3$).

In order to quantify the location in the sky of low metallicity dwarfs we compute their correlation function, weighted by the star formation history. Thus we compute the so-called mark correlation, using star formation as a mark. As can be seen in Fig.~\ref{fig:corr}, where the notation WW/DD indicates that the mark statistic is the ratio of weighted pair counts to unweighted pair counts (in this notation, the traditional unweighted correlation function would be DD/RR, where RR is the number of un--weighted pair counts in a random distribution.). The low metallicity dwarfs  are less correlated than the mean population at scales smaller than $20$ Mpc. The shaded region shows the $2 \sigma$ confidence region from all low metallicity dwarfs  in the SDSS. Our prediction then is that GRBs are to be found in under-dense regions of the dark matter density field, i.e. they will be anti-biased. Most GRBs inhabit regions that show the lowest rates of merging and are undisturbed, despite evidence that some systems do show signs of interaction \citep[e.g.,][]{Fruchter}.

It is curious to point out that planet formation requires exactly the opposite environment: high-metallicity. Therefore, (long) GRBs overall seem to inhabit regions where no planets form, thus  presenting no risk of life extinction. While this is generically correct all over the universe, it is particularly relevant concerning life extictions on Earth. The Milky Way is a rather massive galaxy with very few low metallicity regions. In fact, using the star formation and 
metallicity histories of Milky Way type galaxies \citep{pantermet} in the SDSS sample we find that only 2\% of the Milky Way has metallicity below of $1/10$ of  solar, which we found as a typical upper limit on the metallicity of galaxies that host GRBs. This implies that the expected rate of GRBs in the Milky Way is much than the one simply expected from an estimate based on the average current  cosmic GRB rate per unit time and unit volume ($\sim1.3 {\rm Gpc}^{-3} {\rm yr}^{-1}$). 

\section*{Acknowledgments} 
\label{ack}
We thank the anonymous referee for useful comments on the manuscript. RJ thanks Paolo Padoan for discussions on star formation theory. 
RJ and TP are  most grateful to Annalisa de Cia for many useful suggestions and discussions on a previous version of this manuscript. RJ and TP also thank the Einstein cafe in Bern for hospitality while part of this work was done.
TP  acknowledges support from an ERC advanced grant: GRB.   
Funding for the SDSS and SDSS-II has been provided by the Alfred
P. Sloan Foundation, the Participating Institutions, the
National Science Foundation, the U.S. Department of
Energy, the National Aeronautics and Space Administration,
the Japanese Monbukagakusho, the Max Planck
Society, and the Higher Education Funding Council for
England. The SDSS Web Site is http://www.sdss.org/.


\end{document}